\documentclass[10pt,letterpaper]{article}
\usepackage{opex3}

\begin{document}

\title{ Lasing and suppressed cavity-pulling effect of Cesium active optical clock}

\author{Zhichao Xu, Wei Zhuang,$^*$ and Jingbiao Chen$^*$}

\address{State Key Laboratory of Advanced Optical Communication Systems and Networks, \\Institute of Quantum Electronics, School of Electronics Engineering $\&$ Computer Science, \\Peking University, Beijing 100871, China}

\email{$^*$wzhuang@pku.edu.cn}\email{$^*$jbchen@pku.edu.cn}


\begin{abstract}
We experimentally demonstrate the collective emission behavior and suppressed cavity-pulling effect of four-level active
optical clock with Cesium atoms. Thermal Cesium atoms in a glass
cell velocity selective pumped with a 455.5 nm laser operating at
6S$_{1/2}$ to 7P$_{3/2}$ transition are used as lasing medium.
Population inverted Cesium atoms between 7S$_{1/2}$ and 6P$_{3/2}$
levels are optical weakly coupled by a pair cavity mirrors working
at deep bad-cavity regime with a finesse of 4.3, and the ratio between cavity bandwidth and gain bandwidth is approximately 45. With increased
455.5 nm pumping laser intensity, the output power of cesium active optical clock at 1469.9 nm from 7S$_{1/2}$ level to 6P$_{3/2}$ level shows a threshold
and reach a power of 13 $\mu$W. Active optical clock would
dramatically improve the optical clock stability since the lasing
frequency does not follow the cavity length variation exactly, but
in a form of suppressed cavity pulling effect. In this letter the cavity pulling effect is measured using a Fabry-Perot interferometer (FPI) to be reduced by a factor of 38.2 and 41.4 as the detuning between the 1469.9 nm
cavity length of the Cs active optical clock and the Cs 1469.9 nm transition is set to be
140.8 MHz and 281.6 MHz respectively. The mechanism
demonstrated here is of great significance for new generation optical clocks and can be applied to improve the stability of best optical clocks by at least two orders of magnitude.
\end{abstract}




\section{Introduction}
The optical clocks recently have made great progresses with single
ion~\cite{Chou, Huntemann, Pierre, Gao, Margolis} and neutral atoms
in magic optical
lattice~\cite{Takamoto,Nicholson,Middelmann,McFerran, Hinkly, Bloom}, and their uncertainty and stability of 10$^{-18}$ level
have been realized. However, to improve the clock linewidth to
millihertz level~\cite{Yu1}, is still a huge challenge, which is
currently limited by the thermal Brownian-motion induced cavity
noise of prerequisite stable probing laser used as local oscillator
for all passive optical clocks~\cite{Drever, Young, Jiang, Kessler}.

Up to date all the passive optical clocks are based on probing laser prestabilized to a high-finesse cavity with the Pound-Drever-Hall
(PDH) technique~\cite{Drever, Young, Jiang, Kessler}. The newly reported passive optical clocks with 10$^{-18}$ level uncertainty and stability employ such PDH probing laser systems to interrogate the clock transition spectroscopic line of 6.0 Hz ~\cite{Hinkly} and 5.5 Hz linewidth~\cite{Bloom}. However, the frequency of the PDH probing laser system is sensitive to the cavity-length noise. So far, super-cavities made from ultralow expansion (ULE) glass are used to reduce thermal drift of the cavity-length and
a linewidth of 250 mHz was realized near room temperature~\cite{Jiang}. A sub-40-mHz laser linewidth was realized based on a silicon single-crystal optical cavity operated at 124 K~\cite{Kessler}. To realize narrower PDH laser linewidth, it is necessary to further reduce the optical cavity operating temperature, which is very complicated for optical clocks.

The bad-cavity collective emission scheme with free-perturbation clock transition as laser medium, is proposed as active optical clock~\cite{Kuppens-1,Chen,
Zhuang, Zhuang1, Zhuang2, Chen1, Yu2, Chen2, Wang, Meiser, Sterr,
Meiser1, Meiser2, Yu3, Yu4, Xie, Zhuang3, Zhuang4, Zhuang5, Li,
Bohnet, Bohnet1, Xue, Zang, Kazakov, Bohnet2,Bohnet3,Zhang, Shengnan,
Yanfei, Xu}. Population inverted atoms are prepared in a cavity with much wider cavity mode linewidth than clock transition gain bandwidth. The output of active optical clock is the coherent emission in phase due to weak interaction between collective atoms and cavity field, thus provides much narrower laser linewidth than that of the PDH frequency stabilized laser~\cite{Drever, Young, Jiang, Kessler}. Unlike that in passive optical clock, the self-sustained lasing oscillation of clock transition in bad cavity, instead of the local oscillator frequency stabilized to the clock transition, is used as optical frequency standard in active optical clock. As the cavity bandwidth of active optical clock is designed to be much wider than the clock transition gain bandwidth, i.e. operating deep in the bad cavity regime, the stimulated emission frequency of active optical clock is mainly determined by the free-perturbation clock transition and the influence of cavity mode is sufficiently suppressed, which is quite different from that in traditional lasers working in the good cavity regime.

Since the
proposal of active optical clock~\cite{Chen, Chen2}, a number of
neutral atoms with two-level, three-level, four-level at thermal and
laser cooled and trapped configurations, Raman laser, and sequential
coupling configuration have been investigated recently~\cite{Chen,
Zhuang, Zhuang1, Zhuang2, Chen1, Yu2, Chen2, Wang, Meiser, Sterr,
Meiser1, Meiser2, Yu3, Yu4, Xie, Zhuang3, Zhuang4, Zhuang5, Li,
Bohnet, Bohnet1, Xue, Zang, Kazakov, Bohnet2,Bohnet3,Zhang, Shengnan,
Yanfei, Xu}. The potential quantum limited linewidth of the active
optical clock~\cite{Yu2} is narrower than mHz, and to reach this
unprecedented linewidth is possible since the effect of thermal
noise on cavity mode can be suppressed dramatically with the
mechanism of active optical clock~\cite{Chen, Chen2}. The optical lattice laser based on trapped atoms has been proposed and investigated in \cite{Chen,Meiser,Meiser1} and the limited atomic trap lifetime is the main limitation to its high performance, which can be solved by sequential coupling technique\cite{Kazakov}. The influence of finite atomic-field interaction time on the laser linewidth is theoretically discussed in \cite{Yu2} and active optical clock based on a atomic beam has been studied in\cite{Yu2,Chen2,Zhuang5,Li,Zhang}. Active optical clock based on a thermal $^{88}$Sr atomic beam is proposed in \cite{Chen2} and a 0.5 Hz linewidth with 120 nW power is expected to be realized. An atomic flux of 4.3$\times10^{11}/s$ is needed for self-sustained lasing oscillation. In addition, the bad-cavity
Raman laser configuration has been intensively investigated with
cooled Rb atoms recently~\cite{Meiser1, Meiser2, Bohnet, Bohnet1,
Bohnet2,Bohnet3} with very beautiful results~\cite{Bohnet}. However, for
active optical clocks with laser cooled and trapped atoms in 3-level
configuration~\cite{Chen, Chen1, Meiser, Meiser2, Bohnet, Bohnet1,
Bohnet2, Bohnet3}, the light shift caused by pumping laser is a main
limitation for high performance.

\begin{figure}[htbp]
\centering\includegraphics[width=12cm]{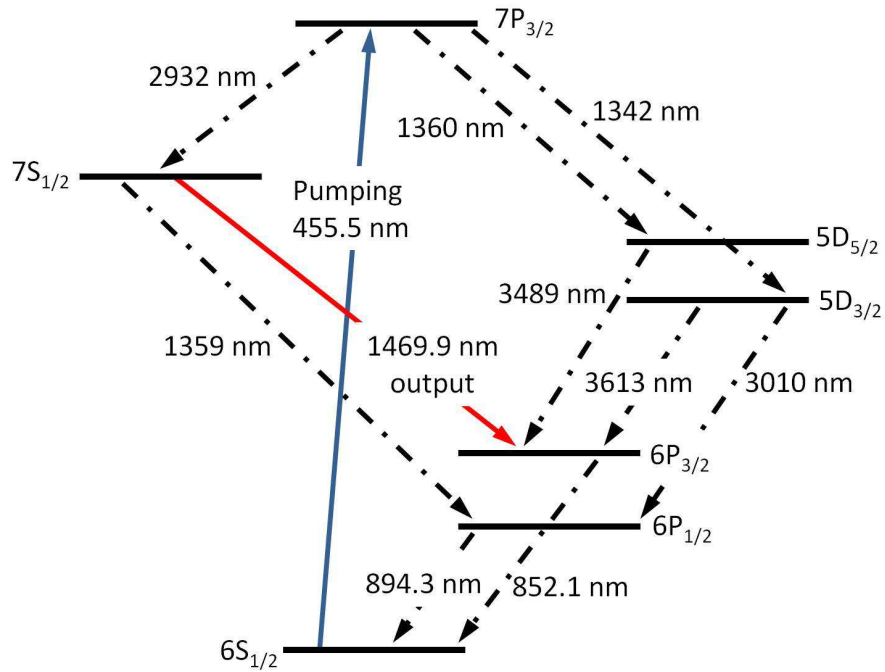}
\caption{(color online). Relevant atomic energy levels of Cs.}
\label{Figure1}
\end{figure}

To avoid this problem, we choose 4-level configuration instead of 3-level configuration. The detuning between the pumping laser frequency and the transitions related to energy levels of clock transition is thus increased from MHz level in 3-level configuration to THz level in 4-level configuration, and the light shift caused by pumping laser can be avoided as the pumping-induced shift is inversely proportional to the detuning. Active optical clock based on four-level quantum system has been investigated\cite{Xue,Zang,Zhang,Shengnan,Yanfei,Xu,Yanfei2,Yanfei3} and the population inversion between 7S$_{1/2}$ and 6P$_{3/2}$ levels of cesium has been experimentally realized\cite{Yanfei} for the lifetime of 7S$_{1/2}$ state is longer than that of 6P$_{3/2}$ state. In this letter, we experimentally demonstrated the stimulated collective
emission behavior and suppressed cavity-pulling effect, the main characteristic of active
optical clock with Cs atoms in four-level configuration. The Cs atoms are prepared in a heated glass cell and thus overcome the atom trap lifetime limitation in active optical clock based on trapped atoms. What's more, the output power of active optical clock based on a heated cell is expected to be much larger than that based on a thermal atomic beam, for there are far more atoms in the heated glass cell than that in the atomic beam. As pointed
previously~\cite{Xue,Zang, Zhang, Shengnan,  Yanfei, Xu, Yanfei2,
Yanfei3}, the energy level structure of alkali metals are
suitable for four-level active optical clock. For Cs, the relevant
atomic energy levels are showed in Fig.1, in which Cs atoms are
excited to 7P$_{3/2}$ from the ground state 6S$_{1/2}$ by 455.5 nm
pumping laser~\cite{Yanfei2, Yanfei3, Dongying}. The 1469.9 nm
output of active optical clock is stimulated emission of radiation
built up between 7S$_{1/2}$ and 6P$_{3/2}$ with weak optical
feedback from cavity.

\section{Experimental schematics}

\begin{figure}[htbp]
\centering\includegraphics[width=12cm]{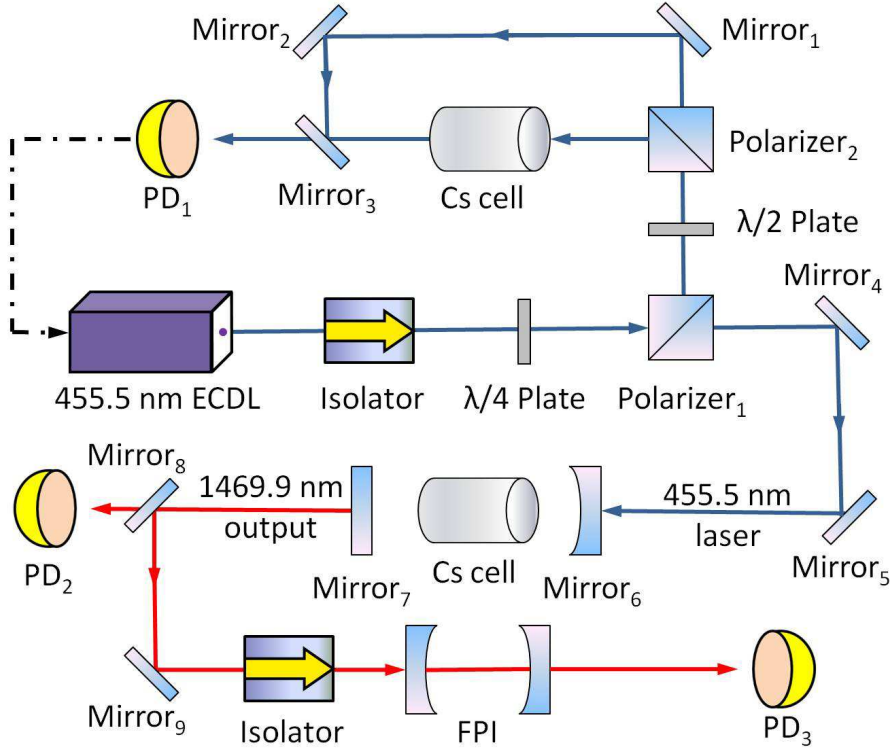}\\
\caption{(color online). Experimental setup of Cs active optical
clock. Mirror$_{1}$, Mirror$_{2}$, Mirror$_{4}$ and Mirror$_{5}$ are
highly reflecting mirrors at 455.5 nm while Mirror$_{3}$ is a
partially reflecting mirror at 455.5 nm. Mirror$_{6}$ is a concave
mirror coated with 455.5 nm anti-reflection and 1469.9 nm
high-reflection coating. Mirror$_{7}$ and Mirror$_{8}$ are plane mirrors coated with 455.5 nm anti-reflection coating
and the reflectivity at 1469.9 nm is 77$\%$, while Mirror$_{9}$ is a highly reflecting mirror at 1469.9 nm.The photodiodes are represented by PD while the Fabry-Perot interferometer is represented by FPI.}\label{Figure2}
\end{figure}

The experimental setup of Cs four-level active optical clock is
showed in Fig.2. The velocity selective pumping scheme where the 455.5 nm cw pumping laser beam is aligned parallel to the 1469.9 nm cavity mode is employed. The 1469.9 nm cavity is composed of a concave
mirror (Mirror$_{6}$ in Fig.2) coated with 1469.9 nm high-reflection
coating and a plane mirror (Mirror$_{7}$ in Fig.2) whose
reflectivity of 1469.9 nm is R=77\%. The parallel-concave cavity scheme ensures the stability of the 1469.9 nm cavity, while the reflectivity of Mirror$_{7}$ in Fig.2 is designed to balance the output power of 1469.9 nm Cs active optical clock and the cavity bandwidth-gain bandwidth ratio. The finesse of the 1469.9 nm cavity without the Cs cell is calculated to be $\frac{\pi\sqrt{R}}{1-R}=12$. The 1469.9 nm cavity mode of  Cs
active optical clock is calibrated in advance with an available 1529
nm laser, which in wavelength is close to the aimed 1469.9 nm. A Cs cell of 5 cm long is inserted
between these two cavity mirrors and the transmission of the Cesium
cell at 1529 nm is T=79.1\%. The Cesium cell is heated to 129 $^{\circ }C$ at which temperature the output power of 1469.9 nm Cs active optical clock almost reaches its maximum value. The
measured reflectivity difference of cavity mirrors and the transmission difference of the heated Cs cell
between 1469.9 nm and 1529 nm are both less than 2.5$\%$. The expected difference of the finesse of these two wavelengths is then calculated to be less than 12\%. Fig.3 shows the
transmitted power of 1529 nm laser through the 1469.9 nm cavity of
Cs active optical clock when changing the cavity length. The doted
blue curve shows the measured cavity mode without the Cs cell and
the finesse is measured to be 12. The
finesse of the cavity with Cs cell is calculated to be $\frac{\pi\sqrt{RT^2}}{1-RT^2}=4.2$.
The full red curve represents the cavity mode with a heated Cesium
cell inserted between the two cavity mirrors and the finesse of the
1469.9 nm cavity is then measured to be F=4.3. The 1469.9 nm cavity length of Cs active optical clock is controlled with a piezoelectric ceramic transducer (PZT) installed on Mirror$_7$ shown in Fig.2 and the length of the 1469.9 nm cavity does not perfectly follows the variation of PZT voltage linearly. Therefore the red curve in Fig.3 seems to be a little asymmetric. The length of the
1469.9 nm cavity is L=8.6 cm and the free spectral range is $FSR=\frac{c}{2L}=1.744 GHz$, where c is the light speed in vacuum. Then the linewidth of 1469.9 nm cavity mode is $\Gamma_{cavity}=\frac{FSR}{F}=405.6 MHz$. The radius of curvature of Mirror$_{6}$ is designed to be r=8000 mm and the waist radius of 1469.9 nm cavity mode is thus calculated to be $\sqrt{\frac{\lambda}{\pi}}L^{\frac{1}{4}}(r-L)^{\frac{1}{4}}=$0.544 mm. The waist radius of 455.5 nm pumping laser beam at the position of the 1469.9 nm cavity of the Cs active optical clock is measured using the Newport's LBP-2-USB laser beam profiler to be 0.463$\times$0.855 mm.

\section{Experimental results and discussion}

\subsection{Lasing of Cesium active optical clock}

\begin{figure}[htbp]
\centering\includegraphics[height=8cm,width=12cm]{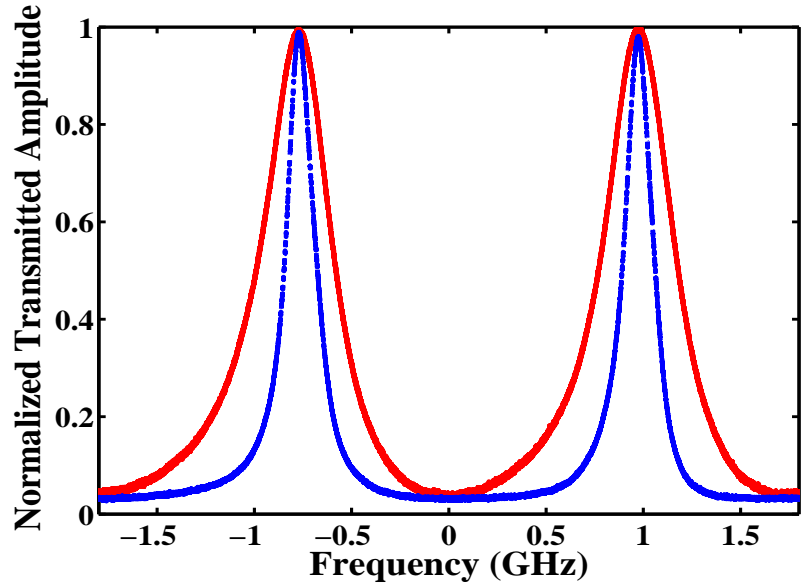}\\
\caption{(color online). The 1469.9 nm cavity modes of Cs active
optical clock calibrated with a 1529 nm laser. The doted blue curve
shows the measured cavity finesse without Cs cell is 12. The full
red curve represents the measured cavity finesse with an inserted Cs
cell is 4.3. The expected difference of the finesse of these two wavelengths is calculated to be less than 12\%.}\label{Figure3}
\end{figure}

\begin{figure}[htbp]
\centering\includegraphics[height=8cm,width=12cm]{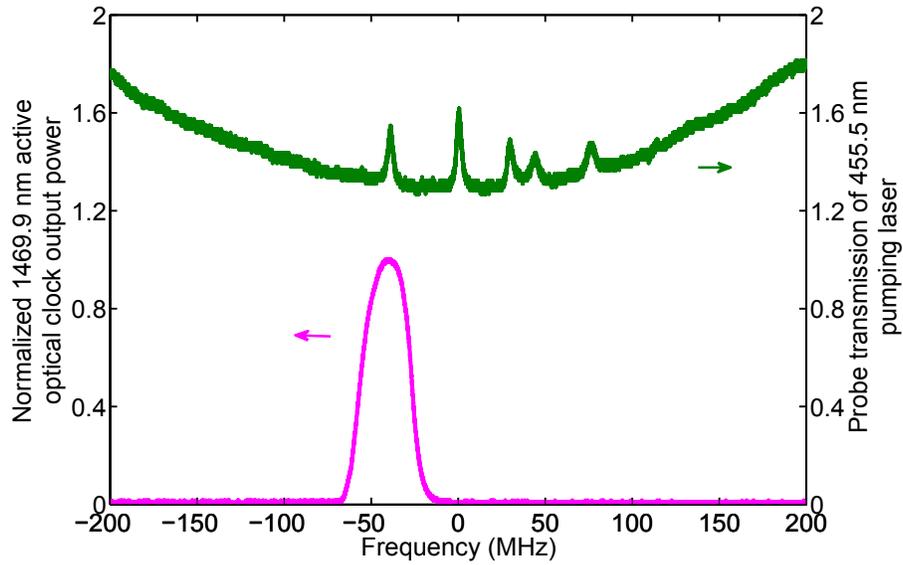}\\
\caption{(color online). The 1469.9 nm output power of Cesium active
optical clock when scanning 455.5 nm pumping laser frequency. The Cs cell is heated to 117
$^{\circ }C$. The upper green curve shows the Cs 6S$_{1/2}$ and
7P$_{3/2}$ saturated absorption spectrum for reference while the lower purple single-peak curve indicates the normalized 1469.9 nm active optical clock output power. The corresponding Y-axises for the upper green curve and the lower purple curve are respectively indicated with the green arrow and the purple arrow.}
\label{Figure4}
\end{figure}

\begin{figure}[htbp]
\centering\includegraphics[height=8cm,width=12cm]{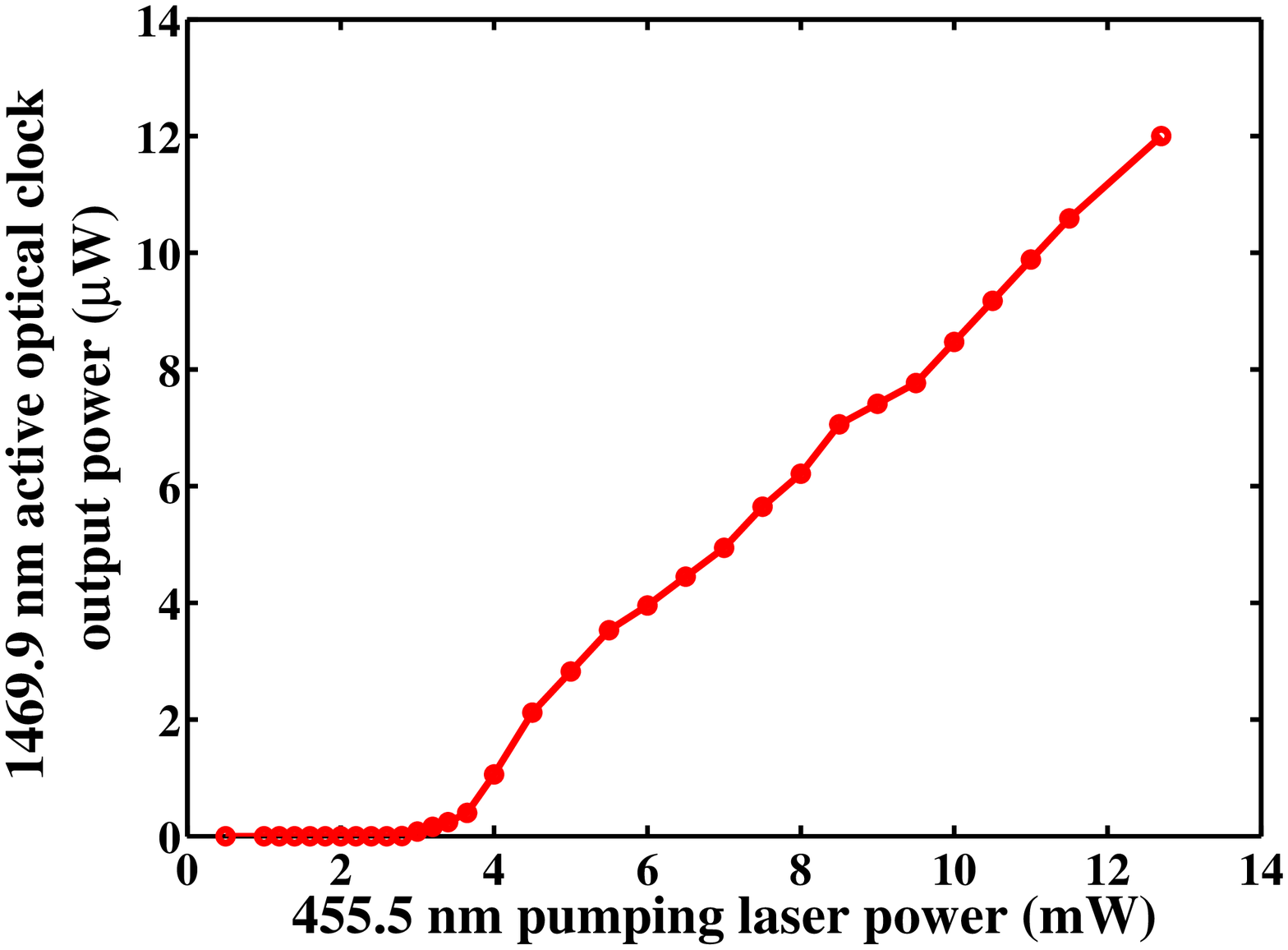}\\
\caption{(color online). Measured 1469.9 nm output power of Cesium
active optical clock when changing 455.5 nm pumping laser power. The
frequency of pumping laser is locked to the cesium 455.5 nm
transition between 6S$_{1/2}$ (F=4) and 7P$_{3/2}$ (F$'$=5) and the
Cesium cell is controlled around 129 $^{\circ }C$. }\label{Figure5}
\end{figure}

Our previous work~\cite{Yanfei, Xu} has showed the population
inversion and lasing between 7S$_{1/2}$ and 6P$_{3/2}$
with 455.5 nm pumping. When the intensity of 455.5 nm pumping laser is
strong enough, the stimulated collective emission of 1469.9 nm radiation will
reach self-sustained lasing oscillation with weak cavity feedback.
It is obvious that the self-sustained 1469.9 nm lasing strongly
depends on the 455.5 nm pumping laser frequency as shown in the
Fig.4, when the power of 455.5 nm pumping laser is 10 mW and the
1469.9 nm cavity length is kept in resonance with the 1469.9 nm Cs
active optical clock output. The 1469.9 nm Cs active optical clock output power (the lower purple trace in Fig.4) is measured while scanning the 455.5 nm pumping laser frequency. Fig.4 shows very clearly the optimized
pumping laser frequency is at 6S$_{1/2}$ (F=4) and 7P$_{3/2}$
(F$'$=5) transition. According to selection rules, only the transition between 7P$_{3/2}$
(F$'$=5) and  7S$_{1/2}$
(F$'$=4) is allowed for the Cs atoms pumped to the 7P$_{3/2}$
(F$'$=5) state. The hyperfine state of 7S$_{1/2}$
(F$'$=4) can thus be proved to be the upper level of 1469.9 nm Cs active optical clock output. The hyperfine state of 6P$_{3/2}$ relevant to the 1469.9 nm output can be determined by population difference between two levels of the clock transition and the transition rate\cite{Reshetov}, which will be reported in our future work.

The measured 1469.9 nm output power of Cs active optical clock, as
showed in Fig.5, increases almost linearly with the 455.5 nm pumping
laser power after a threshold value of 3.1 mW, while the frequency
of pumping laser is locked to the cesium 455.5 nm transition between
6S$_{1/2}$ (F=4) and 7P$_{3/2}$ (F$'$=5) and the Cs cell is
controlled around 129 $^{\circ }C$. When the pumping laser
power is kept to be 11.77 mW, the 1469.9 nm output power of Cs
active optical clock depends on the Cs cell temperature, i.e.,
the atom density. Their relation shows a bell shaped curve as in Fig.6.
Here, the cavity length of Cs active optical clock is kept in position so that
maximum 1469.9 nm
output power is reached. At high Cs
cell temperature, the collisions between dense atoms shorten the
effective coherence time of the 7S$_{1/2}$ excited state in the form of pressure broadening, and cause
the 1469.9 nm output power reduction, even lasing stop. The detailed theoretical explanation has to be further studied and more relevant experimental results are needed.

\begin{figure}[htbp]
\centering\includegraphics[height=8cm,width=12cm]{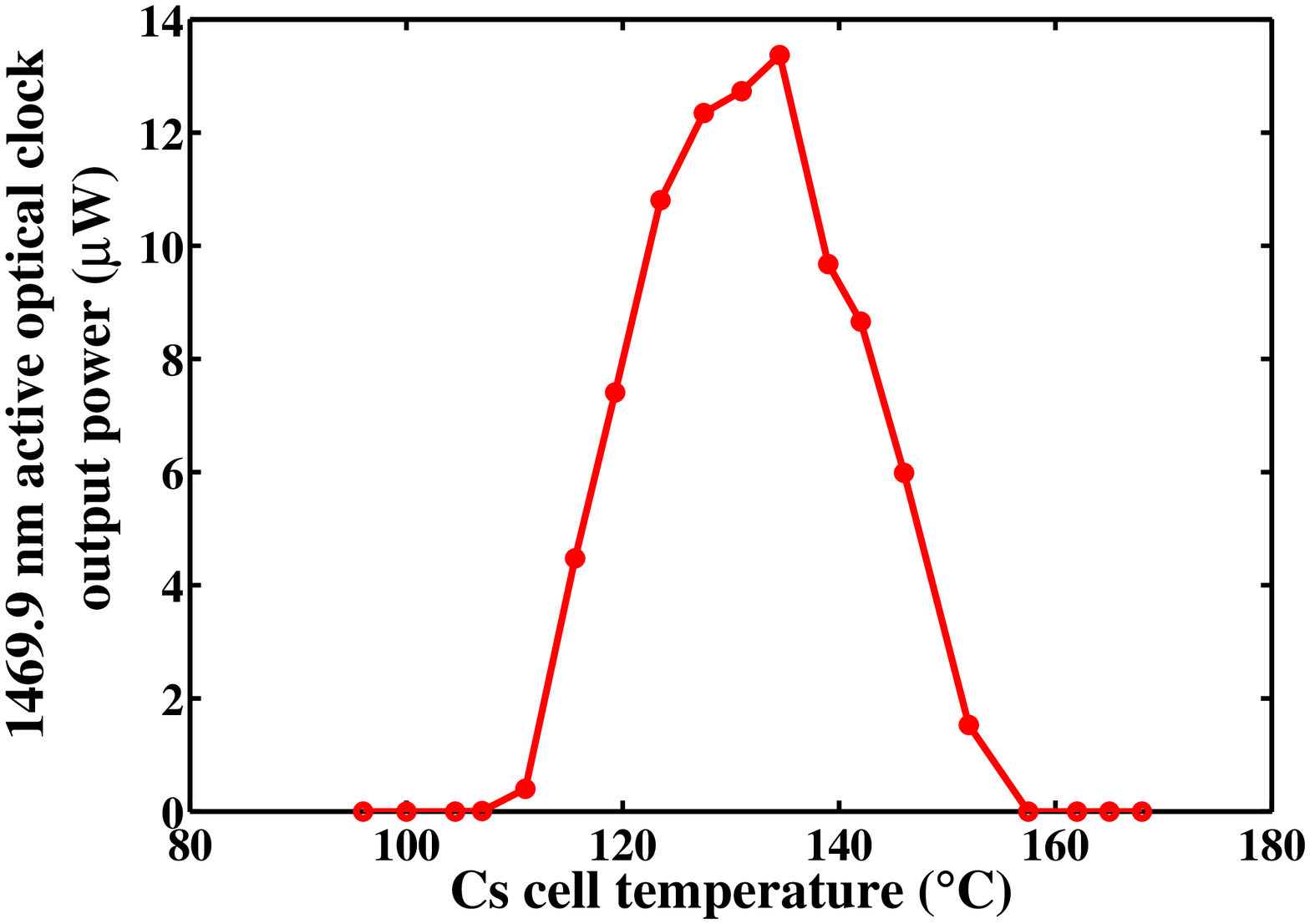}\\
\caption{(color online). Measured output power of 1469.9 nm Cesium active
optical clock when changing the temperature of Cesium cell. The frequency of pumping laser is
locked to the cesium 455.5 nm transition between 6S$_{1/2}$ (F=4)
and 7P$_{3/2}$ (F$'$=5) and the 455.5 nm pumping laser power is kept
to be 11.77 mW.}\label{Figure6}
\end{figure}

\subsection{Suppressed cavity pulling effect of Cesium active optical clock}

\begin{figure}[htbp]
\centering\includegraphics[height=8cm,width=12cm]{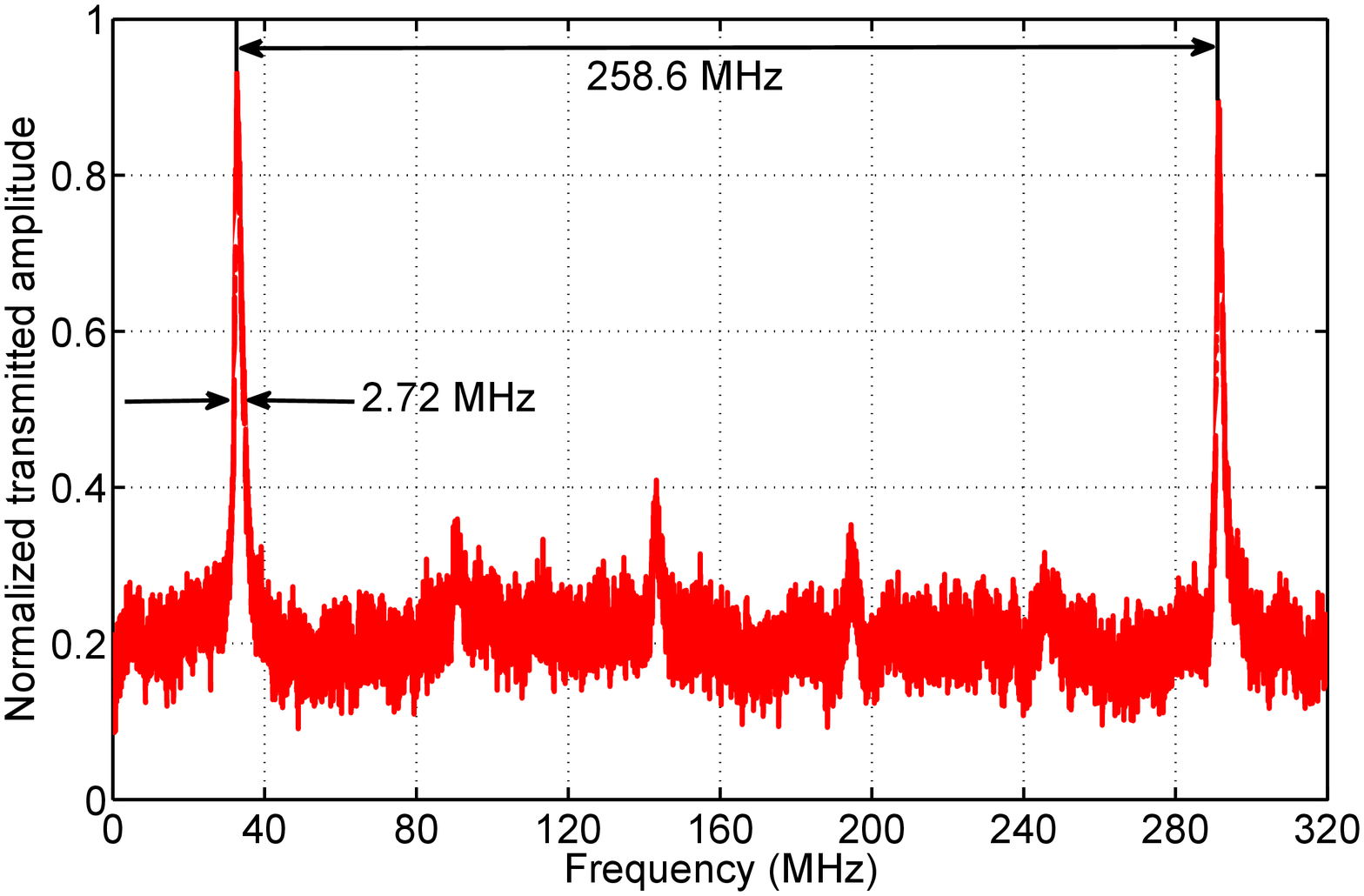}
\caption{ Mode of the 1469.9 nm Cesium active optical clock output measured using a Fabry-Perot
interferometer (FPI) as the setup in Fig.2. The frequency of 11.77 mW
pumping laser is locked to the cesium 455.5 nm transition between
6S$_{1/2}$ (F=4) and 7P$_{3/2}$ (F$'$=5) and the Cs cell temperature
is around 135 $^{\circ }C$.}
\label{feedback2}
\end{figure}

\begin{figure}[htbp]
\centering\includegraphics[height=8cm,width=12cm]{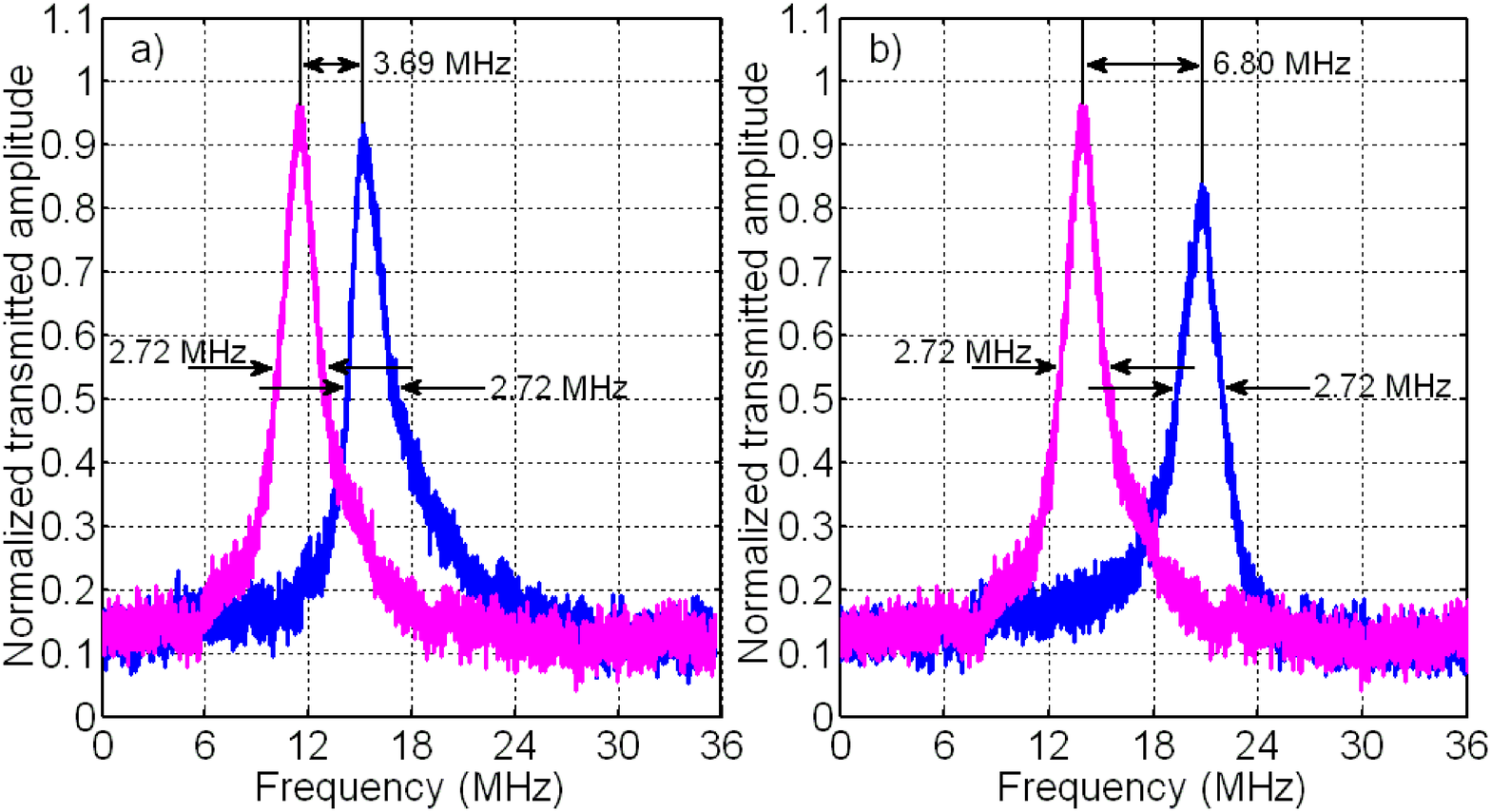}
\caption{ The suppressed cavity-pulling effect in
Cesium active optical clock measured using a Fabry-Perot
interferometer as the setup in Fig.2. The detuning between the 1469.9 nm
cavity length of the Cs active optical clock and the Cs 1469.9 nm transition is set to be a) 140.8 MHz and b) 281.6 MHz.}
\label{feedback1}
\end{figure}

As an active optical clock, the gain bandwidth is designed to be
much narrower than that of cavity mode~\cite{Chen, Chen2}. In this letter the velocity selective pumping scheme is employed, where the 455.5 nm pumping laser beam is aligned parallel to the 1469.9 nm cavity mode of Cs active optical clock and a section of the Doppler profile of atoms in the 455.5 nm pumping laser beam are excited to the 7$P_{3/2}$ state. The corresponding velocity spread parallel to the 1469.9 nm cavity mode can be deduced from the width of the 455.5 nm transition's profile which is determined by the saturation broadening caused by 455.5 nm pumping laser. The
frequency gain bandwidth of 1469.9 nm Cs four-level active optical
clock output is mainly determined by the Doppler broadening of Cs
7S$_{1/2}$ state, which depends on the saturation broadening of Cs
455.5 nm transition via velocity selective pumping. The natural
linewidth of 7P$_{3/2}$ state, the upper level of Cs 455.5 nm
transition, is 1.13 MHz. After considering the power losses of
1469.9 nm cavity mirrors and Cs cell windows, and the measured beam
size of 455.5 nm pumping laser, we estimated the averaged intensity
of 455.5 nm pumping laser within the Cs cell is I=695.2 mW/cm$^2$. Then
the on-resonance saturation parameter is about
S$_{0}$=I/I$_{s}$=434.5 since the saturation intensity of Cs 455.5
nm transition is I$_{s}$=1.60 mW/cm$^{2}$. The broadened Lorentzian
linewidth caused by saturation broadening of Cs 455.5 nm transition, i.e. the width of the 455.5 nm transition's Doppler profile,
is 23.58 MHz and a corresponding velocity spread of 10.74 m/s
parallel to the 1469 9 nm cavity mode of Cs atoms can be pumped to
the 7S$_{1/2}$ state. The Lorentzian linewidth due to the corresponding
Doppler broadening of Cs 7S$_{1/2}$ state is then 7.31 MHz and the
Lorentzian natural linewidth of 1469.9 nm transition is 1.81 MHz. The combined linewidth of two Lorentzian linewidths is the result of their convolution, which equals to the summation of these two linewidths. Thus the total gain bandwidth of 1469.9 nm Cs four-level active
optical clock then is $\Gamma_{gain}=9.12$ MHz while the 1469.9 nm
cavity bandwidth is measured to be $\Gamma_{cavity}$ =405.6 MHz. The
ratio between cavity bandwidth and gain bandwidth
a=$\Gamma_{cavity}$/$\Gamma_{gain}$~\cite{Chen, Chen2}, is a=44.5.

The main characteristic feature and advantage of active optical
clock is the suppressed cavity pulling effect, which shows the
output frequency will not follows the cavity mode changing exactly,
but in a way of dramatically reduced frequency shift when the cavity
mode is changing. The suppressed cavity-pulling effect of 1469.9 nm
output in Cs active optical clock is measured using a Fabry-Perot
interferometer as the FPI in Fig.2. The cavity length of
FPI is 58 cm, thus the free spectral range of FPI
is 258.6 MHz as showed in Fig.7. The finesse of FPI is measured to be 95 and the linewidth of the FPI can thus be calculated to be 2.72 MHz. Fig.7 indicates the single
longitudinal mode normalized transmitted amplitude of 1469.9 nm Cs active
optical clock through FPI when scanning
the FPI cavity length. The additional peaks in Fig.7 are transverse modes of 1469.9 nm Cs active optical clock output and can be sufficiently suppressed by spatially filtering using a aperture slot before the 1469.9 nm is introduced to the FPI. Fig.8 shows the measured
suppressed cavity pulling effect in Cs active optical clock. The
higher red curve represents the 1469.9 nm transmitted signal through the FPI when the 1469.9 nm cavity
length of the Cs active optical clock is kept in resonance with
cesium 1469.9 nm transition, while the lower blue curve represents the 1469.9 nm transmitted signal when the 1469.9 nm cavity
length of the Cs active optical clock is kept 140.8 MHz (Fig.8 (a))and 281.6 MHz (Fig.8 (b))
detuned away from the cesium 1469.9 nm transition. The 140.8 MHz and 281.6 MHz detuning between the 1469.9 nm
cavity mode and atomic 1469.9 nm transition is controlled with
calibrated PZT voltage of 1469.9 nm cavity. By the following
relation\cite{Chen2,Scully},
\begin{equation}
\Delta\nu_{cavity-
pulling}=\frac{1}{1+a}\Delta\nu_{detuning}
\end{equation}
The suppressed cavity pulling $\Delta\nu_{cavity-pulling}$ should only
be 3.10 MHz and 6.20 MHz respectively for 140.8 MHz and 281.6 MHz cavity detuning under current
experimental parameters condition. The averaged value of 15
experimental records of the measured frequency difference due to the
cavity-pulling effect is 3.69 MHz and 6.80 MHz respectively, approximately $1/38.2$ and $1/41.4$ of the
140.8 MHz and 281.6 MHz detuning of cavity away from
the Cs 1469.9 nm transition of active optical clock, and the measured ratio between cavity bandwidth and gain
bandwidth a=37.2 and 40.4 respectively.

The quantum-limited linewidth of the 1469.9 nm Cs active optical clock output can be described by\cite{Kuppens-1,Yu2,Chen2}
\begin{equation}
\Delta\nu=\frac{\Gamma_{cavity}}{2n_{cavity}}\frac{N_p}{{N_p}-{N_s}}(\frac{1}{1+a})^2
\end{equation}
where $N_s$ and $N_p$ represents the populations of the lower and upper levels of Cs 1469.9 nm transition respectively. In the steady state 5.8\% of the total atoms in the cavity mode excited to the 7$P_{3/2}$ state are in 7$S_{1/2}$ state (the upper level) while 2.9\% in the 6$P_{3/2}$ level (the lower level)\cite{Yanfei}. The factor $\frac{N_p}{{N_p}-{N_s}}$ can thus be calculated to be 2. The 1469.9 nm Cs active optical clock output power is measured to be P=13 $\mu$W as shown in Fig.5 and the average photon number in the 1469.9 nm bad cavity of Cs active optical clock is $n_{cavity}=\frac{P}{2{\pi}h\nu\Gamma_{cavity}}=3.8\times10^4$. The quantum--limited linewidth of the 1469.9 nm Cs active optical clock output is then calculated to be 7.3 Hz and 6.2 Hz respectively as the measured ratio between cavity bandwidth and gain
bandwidth a=37.2 and 40.4.

Since the active optical clock output is the stimulated atomic collective emission and the cavity-pulling effect has been sufficiently suppressed, the output frequency of active optical clock is mainly determined by the atomic clock transition instead of the instable macroscopic cavity length. Currently, we are designing a dual wavelength
bad/good cavity, which is coated at 1469.9 nm at bad-cavity regime,
and 632.8 nm at good cavity regime to use a specifically designed He-Ne laser system. The 1469.9 nm laser medium prepared in the heated Cs cell and the 632.8 nm laser medium prepared in the specifically designed He-Ne laser element thus share the same cavity. When the output of good cavity
laser operating at 632.8 nm is locked to a super-cavity with PDH
technique, the stability of 1469.9 nm output at bad-cavity regime will be thus further improved by two orders of magnitude based on the suppressed cavity
pulling effect and the Cesium active optical frequency standard will be then realized. As the optical frequency comb has been a readily available technology, it will be easy to establish the Cesium active optical clock. The mechanism demonstrated here can be applied to other atoms and can be extended to laser cooled and trapped atoms and ions. The research on active optical clock is of great significance for the new generation optical lattice clocks~\cite{Hinkly, Bloom}. We believe the stability of the best optical clock is expected to be improved by at least two orders of magnitude using the mechanism of active optical clock.

\section{Conclusion}

In summary, we have realized the lasing and stimulated collective emission and suppressed cavity-pulling effect of
active optical clock with Cs atoms pumped by a 455.5 nm laser via
velocity selection. The lasing behavior of Cs atoms at 1469.9 nm
radiation was demonstrated while coupled by a bad-cavity with a
finesse of 4.3 only. The characteristics of 455.5 nm pumping laser and cell temperature dependence were exhibited in experiments. Unlike the passive optical clocks where the frequency of local laser oscillator follows the cavity length variation exactly~\cite{Drever, Young, Jiang, Kessler}, the suppressed cavity pulling
effect is the main characteristic of active optical clocks. The four-level active optical
clock could suppress the cavity pulling effect by a factor of 38.2 and 41.4 as the 1469.9 nm cavity
length of the Cs active optical clock is kept 140.8 MHz and 281.6 MHz
detuned away from the cesium 1469.9 nm transition respectively in
this letter, which would dramatically improve stability of optical
clocks. Future experiments will measure the expected narrow
linewidth and demonstrate high performance of active optical clock
by beating and comparing two equal, independent and uncorrelated
experimental setups.

\section*{Acknowledgments}

We thank Chuanwen Zhu for help in 455.5 nm
laser setup, and V. G. Minogin, Longsheng Ma for stimulating discussions. This
work is supported by National Natural Science Foundation of China
under Nos. 10874009 and 11074011, and International Science $\&$ Technology
Cooperation Program of China under No. 2010DFR10900.


\begin{thebibliography}{}


\bibitem{Chou} C. W. Chou, D. B. Hume, J. C. J. Koelemeij, D. J. Wineland, and T. Rosenband, ``Frequency comparison of two high-accuracy Al$^+$ optical clocks," Phys. Rev. Lett. \textbf{104}, 070802 (2010).

\bibitem{Huntemann} N. Huntemann, M. Okhapkin, B. Lipphardt, S. Weyers, Chr. Tamm, and E. Peik, ``High-accuracy optical clock based on the octupole transition in $^{171}$Yb$^+$," Phys. Rev. Lett. \textbf{108}, 090801 (2012).


\bibitem{Pierre} P. Dub$e\acute{}$, A. A. Madej, Z. Zhou, and J.
E. Bernard, ``Evaluation of systematic shifts of the
$^{88}$Sr$^+$
single-ion optical frequency standard at the 10$^{-17}$
level," Phys. Rev. A \textbf{87}, 023806 (2013).

\bibitem{Gao} KeLin Gao, ``Optical frequency standard based on a single $^{40}$Ca$^+$," Chinese Science Bulletin, \textbf{58}, 853-863
(2013).

\bibitem{Margolis} H. S. Margolis, ``Optical frequency standards and clocks," Contemporary Physics \textbf{51} 37-58 (2010).

\bibitem{Takamoto} M. Takamoto, T. Takano, and H. Katori, ``Frequency comparison of optical lattice clocks beyond
the Dick limit,"  Nat. Photonics \textbf{5}, 288-292
(2011).

\bibitem{Nicholson} T. L. Nicholson, M. J. Martin, J. R. Williams, B. J. Bloom, M. Bishof, M. D. Swallows, S. L. Campbell, and J. Ye, ``Comparison of two independent Sr optical clocks with $1\times10^{-17}$ stability at 10$^3$ s,"Phys. Rev. Lett.
\textbf{109}, 230801 (2012).

\bibitem{Middelmann} T. Middelmann, S. Falke, C. Lisdat, and U. Sterr, ``High accuracy correction of blackbody radiation shift in an optical lattice clock," Phys.
Rev. Lett. \textbf{109}, 263004 (2012).

\bibitem{McFerran} J. J. McFerran, L. Yi, S. Mejri, S. Di Manno, W. Zhang, J. Guena, Y. Le Coq, and S. Bize, ``Neutral atom frequency reference in the deep ultraviolet with fractional uncertainty$=5.7\times10^{-15}$," Phys. Rev. Lett. 108, 183004 (2012).
\bibitem{Hinkly} N. Hinkley, J. A. Sherman, N. B. Phillips, M. Schioppo, N. D. Lemke, K. Beloy, M. Pizzocaro, C. W. Oates, and A. D. Ludlow, ``An atomic clock with $10^{-18}$ instability," Science \textbf{341}, 1215-1218 (2013).

\bibitem{Bloom} B. J. Bloom, T. L. Nicholson, J. R. Williams, S. L. Campbell, M. Bishof, X. Zhang, W. Zhang, S. L. Bromley, and J. Ye, ``An optical lattice clock with accuracy and stability at the $10^{-18}$ level," Nature \textbf{000}. 1-5 (2014).


\bibitem{Yu1} D. Yu and J. Chen, ``Optical clock with millihertz linewidth based on a phase-matching effect," Phys. Rev. Lett. \textbf{98}, 050801 (2007).

\bibitem{Drever}  R. W. P. Drever, J. L. Hall, F. V. Kowalski, J. Hough, G. M. Ford, A. J. Munley, and H. Ward, ``Laser phase and frequency stabilization using an optical resonator," Appl. Phys. B \textbf{31}, 97-105 (1983).

\bibitem{Young} B. Young, F. Cruz, W. Itano, J. Bergquist, ``Visible lasers with subhertz linewidths," Phys. Rev. Lett. \textbf{82}, 3799 (1999).

\bibitem{Jiang} Y. Jiang, A. Ludlow, N. Lemke, R. Fox, J. Sherman, L. Ma, and
C. Oates, ``Making optical atomic clocks more stable with $10^{-16}$-level laser stabilization," Nat. Photon. \textbf{5} 158-161 (2011).

\bibitem{Kessler}T. Kessler, C. Hagemann, C.Grebing, T. Legero, U.Sterr, F.Riehle, M.
J. Martin, L.Chen, and J. Ye, ``A sub-40-mHz-linewidth laser based on a silicon single-crystal optical cavity," Nat. Photon. \textbf{6} 687-692 (2012).

\bibitem{Kuppens-1} S. J. M. Kuppens, M. P. van Exter, and J. P. Woerdman , ``Quantum-limited linewidth of a bad-cavity laser," Phys. Rev. Lett. \textbf{72}, 3815 (1994).

\bibitem{Chen} J. Chen and X. Chen, ``Optical lattice laser," \emph{Proceedings of the 2005 IEEE International Frequency Control
Symposium and Exposition}(IEEE, New York, 2005), 608-610.

\bibitem{Zhuang} W. Zhuang, and J. Chen, ``Beyond one-second laser coherence via active optical atomic clock," \emph{Proceedings of the 20th European Frequency and Time
Forum} (Braunschweig, Germany, 2006), 373-375.

\bibitem{Zhuang1} W. Zhuang, D. Yu, and J. Chen, ``Optical clocks based on quantum emitters, " \emph{Proceedings of the 2006 IEEE International Frequency Control Symposium and
Exposition}(Miami, Florida, 2006), 277-280.

\bibitem{Zhuang2} W. Zhuang, D. Yu, Z. Chen, K. Huang, and J. Chen. ``Proposed active optical frequency standards based on magneto-optical trap trapped atoms," \emph{Proceedings of the IEEE Frequency Control Symposium \& European Frequency
and Time Forum} (Geneva,. Switzerland, 2007), 96-99.

\bibitem{Yu2} D. Yu and J. Chen, ``Laser theory with finite atom-field interacting time," Phys. Rev. A \textbf{78}, 013846 (2008).

\bibitem{Chen1} J. Chen, ``Active optical clocks," in \emph{Frequency Standards and
Metrology: Proceedings of the 7th Symposium}, edited by Maleki Lute
(World Scientic, Singapore, 2009), 525-531.

\bibitem{Chen2} J. Chen, ``Active optical clock," Chin. Sci. Bull. \textbf{54}, 348-352 (2009).

\bibitem{Wang} Y. Wang, ``Optical clocks based on stimulated emission radiation," Chin. Sci. Bull. \textbf{54}, 347 (2009).

\bibitem{Meiser} D. Meiser, J. Ye, D. R. Carlson, and M. J. Holland, ``Prospects for a millihertz-linewidth laser," Phys. Rev. Lett. \textbf{102}, 163601 (2009).

\bibitem{Sterr} Uwe Sterr and Christian Lisdat, ``Millihertz-linewidth lasers: A sharper laser," Nature physics \textbf{5}, 382-383 (2009).

\bibitem{Meiser1} D. Meiser and M. J. Holland, ``Steady-state superradiance with alkaline-earth-metal atoms," Phys. Rev. A \textbf{81}, 033847
(2010).

\bibitem{Meiser2} D. Meiser and M. J. Holland, ``Intensity fluctuations in steady-state superradiance," Phys. Rev. A \textbf{81}, 063827 (2010).

\bibitem{Yu3} D. Yu and J. Chen, ``Four-level superradiant laser with full atomic cooperativity," Phys. Rev. A \textbf{81}, 053809 (2010).


\bibitem{Yu4} D. Yu and J. Chen, ``
Theory of quenching quantum fluctuations of a laser system with a ladder-type configuration," Phys. Rev. A \textbf{81}, 023818 (2010).

\bibitem{Xie} X. Xie, W. Zhuang, and J. Chen, ``Adiabatic passage based on the Calcium active optical clock," Chin. Phys. Lett. \textbf{27}, 074202
(2010).

\bibitem{Zhuang3} W. Zhuang and J. Chen, ``Progress of active optical frequency standard based on thermal Ca
atomic beam," \emph{Proceedings of 2010 IEEE International Frequency Control Symposium}(Newport Beach, California, 2010), p. 222-223.

\bibitem{Zhuang4} W. Zhuang, T. Zhang, and J. Chen, ``Active ion optical clock," arXiv:1111.4704 (2011).

\bibitem{Zhuang5} W. Zhuang, J. Chen, ``Feasibility of extreme ultraviolet active optical clock," Chin. Phys. Lett. \textbf{28} 080601 (2011).

\bibitem{Li} Yang Li, Wei Zhuang, Jinbiao Chen, Hong Guo, ``The linewidth of Ramsey laser with bad cavity," arXiv:1001.2670 (2010).

\bibitem{Bohnet} J. G. Bohnet, Z. Chen, J. M. Weiner, D. Meiser, M. J. Holland, and J. K. Thompson, ``A steady-state superradiant laser with less than one intracavity photon," Nature \textbf{484}, 78-81
(2012).

\bibitem{Bohnet1} J. G. Bohnet, Z. Chen, J. M. Weiner, K. C. Cox, and J. K. Thompson, ``Relaxation oscillations, stability, and cavity feedback in a superradiant Raman laser," Phys. Rev. Lett. \textbf{109}, 253602 (2012).

\bibitem{Xue} Y. Wang X. Xue, D. Wang, T. Zhang, Q. Sun, Y. Hong, W. Zhuang, and J. Chen, ``Cesium active optical clock in four-level laser
configuration" in \emph{Proceedings of 2012 IEEE International Frequency Control Symposium}(Baltimore, Maryland, 2012), p. 1-4.

\bibitem{Zang}X. Zang, T. Zhang, and J. Chen, ``Magic wavelengths for a lattice trapped Rubidium four-Level active optical clock," Chin. Phys. Lett. \textbf{29}, 090601 (2012).

\bibitem{Kazakov} G. A. Kazakov, and T. Schumm, ``Active optical frequency standard using sequential coupling of atomic ensembles," Phys. Rev. A  \textbf{87}, 013821 (2013).

\bibitem{Bohnet2} J. G. Bohnet, Z. Chen, J. M. Weiner, K. C. Cox, and J. K. Thompson, ``
Active and passive sensing of collective atomic coherence in a superradiant laser," Phys. Rev. A \textbf{88}, 013826
(2013).

\bibitem{Bohnet3} J. G. Bohnet, Z. Chen, J. M. Weiner, K. C. Cox, and J. K. Thompson, ``
Linear-response theory for superradiant lasers," Phys. Rev. A \textbf{89}, 013806
(2014).

\bibitem{Zhang} T. Zhang, Y. Wang, X. Zang, W. Zhuang, and J. Chen, ``Active optical clock based on four-level quantum system," Chin. Sci. Bull. \textbf{58},
2033-2038 (2013).

\bibitem{Shengnan} S. Zhang, Y. Wang, T. Zhang, W. Zhuang and J. Chen, ``A Potassium atom four-Level active optical clock scheme," Chin. Phys. Lett.
\textbf{30}, 040601 (2013).

\bibitem{Yanfei} Y. Wang, D. Wang, T. Zhang, Y. Hong, S. Zhang, Z. Tao, X. Xie, and J. Chen, ``Realization of population inversion between 7S$_{1/2}$ and 6P$_{3/2}$ levels of cesium for four-level active optical clock," Science China, Physics, Mechanics and Astronomy  \textbf{56}, 1107-1110 (2013).
\bibitem{Xu} Z. Xu, W. Zhuang, Y. Wang, D. Wang, X. Zhang, X. Xue, D. Pan, and J. Chen, ``Lasing of Cesium four-level active optical clock," \emph{Joint IEEE International Frequency Control Symposium \& European Frequency and Time Forum}(Prague,
Czech Repblic, 2013), p. 395-398.

\bibitem{Yanfei2} Y. Wang, S. Zhang, D. Wang, Z. Tao, Y. Hong, and J.Chen, ``Nonlinear optical filter with ultranarrow bandwidth approaching the natural linewidth," Opt. Lett. \textbf{37}, 4059-4061 (2012).
\bibitem{Yanfei3} Y. Wang, X. Zhang, D. Wang, Z. Tao, W. Zhuang, and J. Chen, ``Cs Faraday optical filter with a single transmission peak resonant with the atomic transition at 455 nm," Opt. Express \textbf{20}, 25817-25825 (2012).

\bibitem{Dongying} D. Wang, Y. Wang, Z. Tao, S. Zhang, Y. Hong, W. Zhuang, and J. Chen, ``Cs 455 nm nonlinear spectroscopy with ultra-narrow linewidth," Chin. Phys. Lett. \textbf{30}, 060601 (2013).

\bibitem{Reshetov} V. A. Reshetov, ``
Polarization properties of superradiance from levels with hyperfine structure," J. Phys. B \textbf{28}, p. 1899-1904.
(2014).

\bibitem{Scully} M. O. Scully, and M. S. Zubairy, \emph{Quantum Optics}, (Cambridge University Press, New York, NY, 1997).



\end{thebibliography}
\end{document}